\documentclass{rmf-d}
\usepackage{nopageno,rmfbib,multicol,times,epsf,amsmath,amssymb,cite}
\usepackage[T1]{fontenc} 
\usepackage[]{caption2}
\usepackage{graphicx}
\usepackage{blindtext}
\usepackage{color}
\usepackage{hyperref}
\usepackage{adjustbox}
\def\rmfcornisa{Research OR Education of Physics \hfill\rmf\ {\bf ??} (*?*) ???--???
\hfill MES? A\~NO?}

\def\rmfcintilla{{\it Rev.\ Mex.\ Fis.\/} {\bf ??} (*?*) (????) ???--???}
\clearpage \rmfcaptionstyle 
\begin{document}
\markboth{ RMF Editorial Team    }{ A \LaTeX template for the RMF, RMF-E, SRMF }

%

\newcommand{\pp}           {pp\xspace}
\newcommand{\ppbar}        {\mbox{$\mathrm {p\overline{p}}$}\xspace}
\newcommand{\XeXe}         {\mbox{Xe--Xe}\xspace}
\newcommand{\PbPb}         {\mbox{Pb--Pb}\xspace}
\newcommand{\pA}           {\mbox{pA}\xspace}
\newcommand{\pPb}          {\mbox{p--Pb}\xspace}
\newcommand{\AuAu}         {\mbox{Au--Au}\xspace}
\newcommand{\dAu}          {\mbox{d--Au}\xspace}

\newcommand{\s}            {\ensuremath{\sqrt{s}}\xspace}
\newcommand{\sNN}          {\ensuremath{\sqrt{s_{\mathrm{NN}}}}\xspace}
\newcommand{\pT}           {\ensuremath{p_{\rm T}}\xspace}
\newcommand{\meanpt}       {$\langle p_{\mathrm{T}}\rangle$\xspace}
\newcommand{\ycms}         {\ensuremath{y_{\rm CMS}}\xspace}
\newcommand{\ylab}         {\ensuremath{y_{\rm lab}}\xspace}
\newcommand{\etarange}[1]  {\mbox{$\left | \eta \right |~<~#1$}}
\newcommand{\yrange}[1]    {\mbox{$\left | y \right |~<~#1$}}
\newcommand{\dndy}         {\ensuremath{\mathrm{d}N_\mathrm{ch}/\mathrm{d}y}\xspace}
\newcommand{\dndeta}       {\ensuremath{\mathrm{d}N_\mathrm{ch}/\mathrm{d}\eta}\xspace}
\newcommand{\avdndeta}     {\ensuremath{\langle\dndeta\rangle}\xspace}
\newcommand{\dNdy}         {\ensuremath{\mathrm{d}N_\mathrm{ch}/\mathrm{d}y}\xspace}
\newcommand{\Npart}        {\ensuremath{N_\mathrm{part}}\xspace}
\newcommand{\Ncoll}        {\ensuremath{N_\mathrm{coll}}\xspace}
\newcommand{\dEdx}         {\ensuremath{\textrm{d}E/\textrm{d}x}\xspace}
\newcommand{\RpPb}         {\ensuremath{R_{\rm pPb}}\xspace}

\newcommand{\nineH}        {$\sqrt{s}~=~0.9$~Te\kern-.1emV\xspace}
\newcommand{\seven}        {$\sqrt{s}~=~7$~Te\kern-.1emV\xspace}
\newcommand{\twoH}         {$\sqrt{s}~=~0.2$~Te\kern-.1emV\xspace}
\newcommand{\twosevensix}  {$\sqrt{s}~=~2.76$~Te\kern-.1emV\xspace}
\newcommand{\five}         {$\sqrt{s}~=~5.02$~Te\kern-.1emV\xspace}
\newcommand{\twosevensixnn}{$\sqrt{s_{\mathrm{NN}}}~=~2.76$~Te\kern-.1emV\xspace}
\newcommand{\fivenn}       {$\sqrt{s_{\mathrm{NN}}}~=~5.02$~Te\kern-.1emV\xspace}
\newcommand{\fivennFour}       {$\sqrt{s_{\mathrm{NN}}}~=~5.44$~Te\kern-.1emV\xspace}
\newcommand{\fivennThree}       {$\sqrt{s_{\mathrm{NN}}}~=~5.36$~Te\kern-.1emV\xspace}

\newcommand{\LT}           {L{\'e}vy-Tsallis\xspace}
\newcommand{\GeVc}         {Ge\kern-.1emV/$c$\xspace}
\newcommand{\MeVc}         {Me\kern-.1emV/$c$\xspace}
\newcommand{\TeV}          {Te\kern-.1emV\xspace}
\newcommand{\GeV}          {Ge\kern-.1emV\xspace}
\newcommand{\MeV}          {Me\kern-.1emV\xspace}
\newcommand{\GeVmass}      {Ge\kern-.2emV/$c^2$\xspace}
\newcommand{\MeVmass}      {Me\kern-.2emV/$c^2$\xspace}
\newcommand{\lumi}         {\ensuremath{\mathcal{L}}\xspace}

\newcommand{\ITS}          {\rm{ITS}\xspace}
\newcommand{\TOF}          {\rm{TOF}\xspace}
\newcommand{\ZDC}          {\rm{ZDC}\xspace}
\newcommand{\ZDCs}         {\rm{ZDCs}\xspace}
\newcommand{\ZNA}          {\rm{ZNA}\xspace}
\newcommand{\ZNC}          {\rm{ZNC}\xspace}
\newcommand{\SPD}          {\rm{SPD}\xspace}
\newcommand{\SDD}          {\rm{SDD}\xspace}
\newcommand{\SSD}          {\rm{SSD}\xspace}
\newcommand{\TPC}          {\rm{TPC}\xspace}
\newcommand{\TRD}          {\rm{TRD}\xspace}
\newcommand{\VZERO}        {\rm{V0}\xspace}
\newcommand{\VZEROA}       {\rm{V0A}\xspace}
\newcommand{\VZEROC}       {\rm{V0C}\xspace}
\newcommand{\Vdecay} 	   {\ensuremath{V^{0}}\xspace}

\newcommand{\ee}           {\ensuremath{e^{+}e^{-}}} 
\newcommand{\pip}          {\ensuremath{\pi^{+}}\xspace}
\newcommand{\pim}          {\ensuremath{\pi^{-}}\xspace}
\newcommand{\kap}          {\ensuremath{\rm{K}^{+}}\xspace}
\newcommand{\kam}          {\ensuremath{\rm{K}^{-}}\xspace}
\newcommand{\pbar}         {\ensuremath{\rm\overline{p}}\xspace}
\newcommand{\kzero}        {\ensuremath{{\rm K}^{0}_{\rm{S}}}\xspace}
\newcommand{\lmb}          {\ensuremath{\Lambda}\xspace}
\newcommand{\almb}         {\ensuremath{\overline{\Lambda}}\xspace}
\newcommand{\Om}           {\ensuremath{\Omega^-}\xspace}
\newcommand{\Mo}           {\ensuremath{\overline{\Omega}^+}\xspace}
\newcommand{\X}            {\ensuremath{\Xi^-}\xspace}
\newcommand{\Ix}           {\ensuremath{\overline{\Xi}^+}\xspace}
\newcommand{\Xis}          {\ensuremath{\Xi^{\pm}}\xspace}
\newcommand{\Oms}          {\ensuremath{\Omega^{\pm}}\xspace}
\newcommand{\degree}       {\ensuremath{^{\rm o}}\xspace}

\newcommand{\sF}{\s~=~5.02~TeV\xspace}
\newcommand{\sT}{\s~=~13~TeV\xspace}
\newcommand{\piz}{\ensuremath{\pi^{0}}\xspace}
\newcommand{\e}{\ensuremath{\eta}\xspace}
\newcommand{\Vo}{V\ensuremath{^{0}}\xspace}

\newcommand{\Ngchrec}{\ensuremath{N_{\mathrm{\gamma}}^{\mathrm{\tiny rec}}/ N_{\mathrm{\tiny ch}}^{\mathrm{\tiny rec}}}\xspace}

\newcommand{\Psipairm}{\ensuremath{\Psi_\text{\tiny Pair}}}
\newcommand{\Ng}{\ensuremath{N_{\rm{\gamma}}}\xspace}
\newcommand{\Ngrec}{\ensuremath{N_{\rm{\gamma}}^{\rm{\tiny rec}}}\xspace}
\newcommand{\Ngreci}{\ensuremath{N_{{\gamma,i}}^{\rm{\tiny rec}}}\xspace}
\newcommand{\NgreciX}{\ensuremath{N_{{\gamma,i}}^{\rm{\tiny rec,X}}}\xspace}
\newcommand{\NgreciMC}{\ensuremath{N_{{\gamma,i}}^{\rm{\tiny rec,MC}}}\xspace}

\newcommand{\Ngprod}{\ensuremath{N_{\rm{\gamma}}^{\rm{\tiny prod}}}\xspace}
\newcommand{\Nch}{\ensuremath{N_{\rm{\tiny {ch}}}}\xspace}
\newcommand{\Nchprod}{\ensuremath{N_{\rm{\tiny {ch}}}^{\rm{\tiny prod}}}\xspace}

\newcommand{\Nchrec}{\ensuremath{N_{\rm{\tiny {ch}}}^{\rm{\tiny rec}}}\xspace}

\newcommand{\NchrecX}{\ensuremath{N_{\rm{\tiny {ch}}}^{\rm{\tiny rec,X}}}\xspace}

\newcommand{\Ngch}{\ensuremath{N_{\rm{\gamma}}/N_{\rm{\tiny ch}}}\xspace}

\newcommand{\RConv}{\ensuremath{R_{\rm{\tiny conv}}}}
\newcommand{\ZConv}{\ensuremath{Z_{\rm{\tiny conv}}}}

\newcommand{\qT}           {\ensuremath{q_{\rm T}}\xspace}
\newcommand{\pTmin}           {\ensuremath{p_{\rm T,min}}\xspace}

\newcommand{\Rg}{\ensuremath{R_{\rm{\gamma}}}\xspace}

\newcommand{\mT}           {\ensuremath{m_{\rm T}}\xspace}
\newcommand{\RAA}{\ensuremath{R_{\rm{AA}}}\xspace}
\newcommand{\mee}           {\ensuremath{m_{\rm ee}}\xspace}

%
%
\title{Latest results of the ALICE Collaboration and plans for ALICE 3\vspace{-6pt}}
%
%
\author{A. Marin,  for the ALICE Collaboration}
\address{ GSI Helmholtzzentrum f\"ur Schwerionenforschung GmbH, Planckstrasse 1 , 64291 Darmstadt, Germany}
\maketitle
%
%
\recibido{\today}{DD MM 2023
\vspace{-12pt}}
\begin{abstract}
\vspace{1em} 

%
%
The ALICE experiment is devoted to the study of the quark-gluon plasma (QGP) created in heavy-ion collisions at the CERN LHC. The experimental setup allows for the study of many different observables that contributed
to the characterization of the properties of the QGP. The ALICE experiment went through a major upgrade during LS2 to profit from the increased LHC luminosity and to improve the tracking resolution. An additional upgrade 
is planned for LS3. 
A new experiment,  ALICE 3, was proposed as next major upgrade in LS4. In this contribution, a selection of recent physics results  
was presented together with a  glimpse of the next upgrades during LS3 and LS4, with the main focus on ALICE 3 and its physics program.

\vspace{1em}
\end{abstract}
\keys{ \bf{\textit{QGP, heavy-ion, LHC, ALICE, ALICE 3
}} \vspace{-8pt}}
\begin{multicols}{2}

\section{Introduction}
Heavy-ion collisions at ultrarelativistic energies allow to explore the phase diagram of QCD matter. Lattice QCD calculations predict that a state of matter 
where quarks and gluons are deconfined, known as the QGP, is produced above a given energy density.
In order to characterize the QGP several probes can be used, each of them being sensitive to different properties. Furthermore, the complete evolution of the 
heavy-ion collision is  studied, from the pre-equilibrium phase, the QGP phase, hydrodynamical expansion and hadron gas phase up to the kinetic and thermal freeze-out.

\begin{figure}[H]
 \hspace*{-0.25cm}\includegraphics[width=0.54\linewidth,angle=-90,trim={5cm 2cm 4.5cm 2cm }, clip]{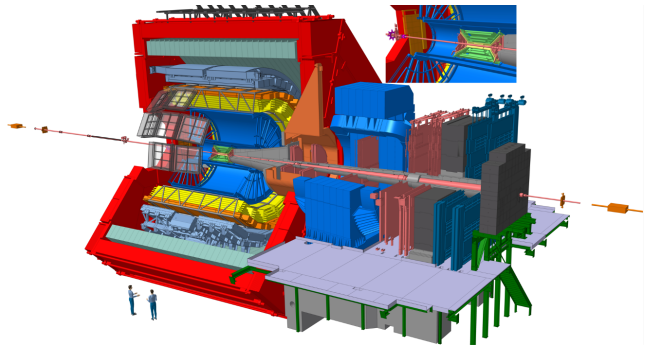}
 \caption{Schematic view of the ALICE apparatus during Run 2.}
 \label{fig:ALICEsetup}
\end{figure}

\begin{figure}[H]
 \includegraphics[width=0.124\linewidth,angle=-90,trim={9cm 0cm 9.cm 0cm }, clip]{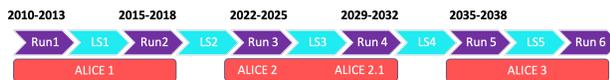}
  \caption{LHC schedule with the different data taking runs, shutdown periods and ALICE configurations.}
 \label{fig:LHCTimeLine}
\end{figure}

Figure {\ref{fig:ALICEsetup}}  displays  the ALICE experiment at the CERN LHC during Run 2. The description of the apparatus and its performance can be found 
in  Refs. \cite{Aamodt:2008zz,Abelev:2014ffa}. The LHC schedule with the different data taking runs, shutdown periods and ALICE configurations is shown in Fig. \ref{fig:LHCTimeLine}.
The ALICE collaboration consists of more than 2000 people distributed over more than 175 institutes  in about 40 countries.  A large number of physics results was published in 
more than 400 papers. A review of the most important results was made available recently \cite{ALICE:2022wpn}. In the following, few selected recent results will be described.
Furthermore, some details on the already started Run 3 are given as well as a glimpse of  the future upgrades and the physics program.
\noindent

\section{Global properties}

The multiplicity of charged particles produced at mid-rapidity in heavy-ion collisions  is a key observable to characterize the properties of the matter
created in these collisions. The  dependence of d$N_{\rm ch}$/d\e on collision system, center-of-mass energy and collision geometry are
 basic observables for understanding the different particle production mechanisms.
ALICE measured  a power law dependence on the charged particle multiplicity density as a function of the  center of mass energy per nucleon pair in \pp and \PbPb collisions \cite{ALICE:2015juo}.
A larger  exponent  for nuclear collisions as compared to \pp collisions evidences that more  energy is available for particle production in these collisions compared to \pp. 

One of the strengths of ALICE is its particle identification capabilities as a result of using most of the available techniques and the low transverse momentum (\pT) reach. 
Thus, a broad set of high precision measurements of identified particles
differentially as a function of transverse momentum and  collision centrality were carried out in \pp, \pPb and \PbPb collisions \cite{ALICE:2019hno}. 
The spectra become flatter with increasing charged particle multiplicity,  the effect being more pronounced for heavier particles.
The low \pT spectra were parametrized  with blast-wave fits across the systems obtaining freeze-out temperature $T_{\rm kin}$ and radial flow velocities $\beta_{\rm T}$ that depend strongly on d$N_{\rm ch}$/d\e \cite{ALICE:2020nkc}.
With all these differential measurements, computed  integrated yields of
different particle species with respect to those of charged pions were studied as a function of charged particle multiplicity with the main focus on strange and multistrange particles \cite{ALICE:2016fzo} as 
strangeness enhancement was proposed as a signature of QGP formation in nuclear collisions. 
A continuous evolution of strangeness production between different collision systems and energies is observed. The magnitude of the strangeness enhancement grows with the strange quark content.
The hadron yields are mostly related to the final state charged particle multiplicity density rather than collision system or beam energy. 
Furthermore, multiplicities of hadron species  containing only light quarks measured at midrapidity in central \PbPb collisions, spanning over 
nine orders of magnitude in abundance values, are well described by statistical hadronization models \cite{ALICE:2022wpn}. 

Studies of the azimuthal anisotropy of particle production have contributed significantly to the characterization of the system created in heavy-ion collisions. 
Anisotropic flow reflects the conversion of the initial state spatial anisotropy into final state anisotropies in momentum space. 
Elliptic flow in \PbPb, \pPb and \pp collision was measured \cite{ALICE:2022zks} using two ($v_2\{2\}$) and four-particle ($v_2\{4\}$) cumulants for different particle species. A mass ordering was found 
at low \pT, and at intermediate \pT baryon vs meson grouping that can be interpreted as  quark-level flow and recombination. Models including quark coalescence describe the measurements over a large \pT range, which confirms the relevance of the quark coalescence hadronization mechanism in the particle production in \pp, \pPb and \PbPb collisions at the LHC.

Viscous hydrodynamic calculations using various initial state models were able to describe multiplicity distributions, particle momentum spectra and integrated flow measurements. Uncertainties in the initial state 
translated into large uncertainties on the extracted shear viscosity over entropy ratio, \e/$s$, value. Some progress was achieved in the last years. 
Novel approaches utilizing Bayesian statistics and a multiparameter model-to-data comparison of different observables like multiplicity, transverse momentum, and flow are used to quantitatively 
extract estimates of the temperature-dependent specific shear and bulk viscosity simultaneously with related initial-condition properties \cite{Bernhard:2019bmu} 
or to characterize additional aspects of high-energy nuclear collisions \cite{JETSCAPE:2020mzn}.

\begin{figure}[H]
\begin{center}
 \includegraphics[width=0.7\linewidth]{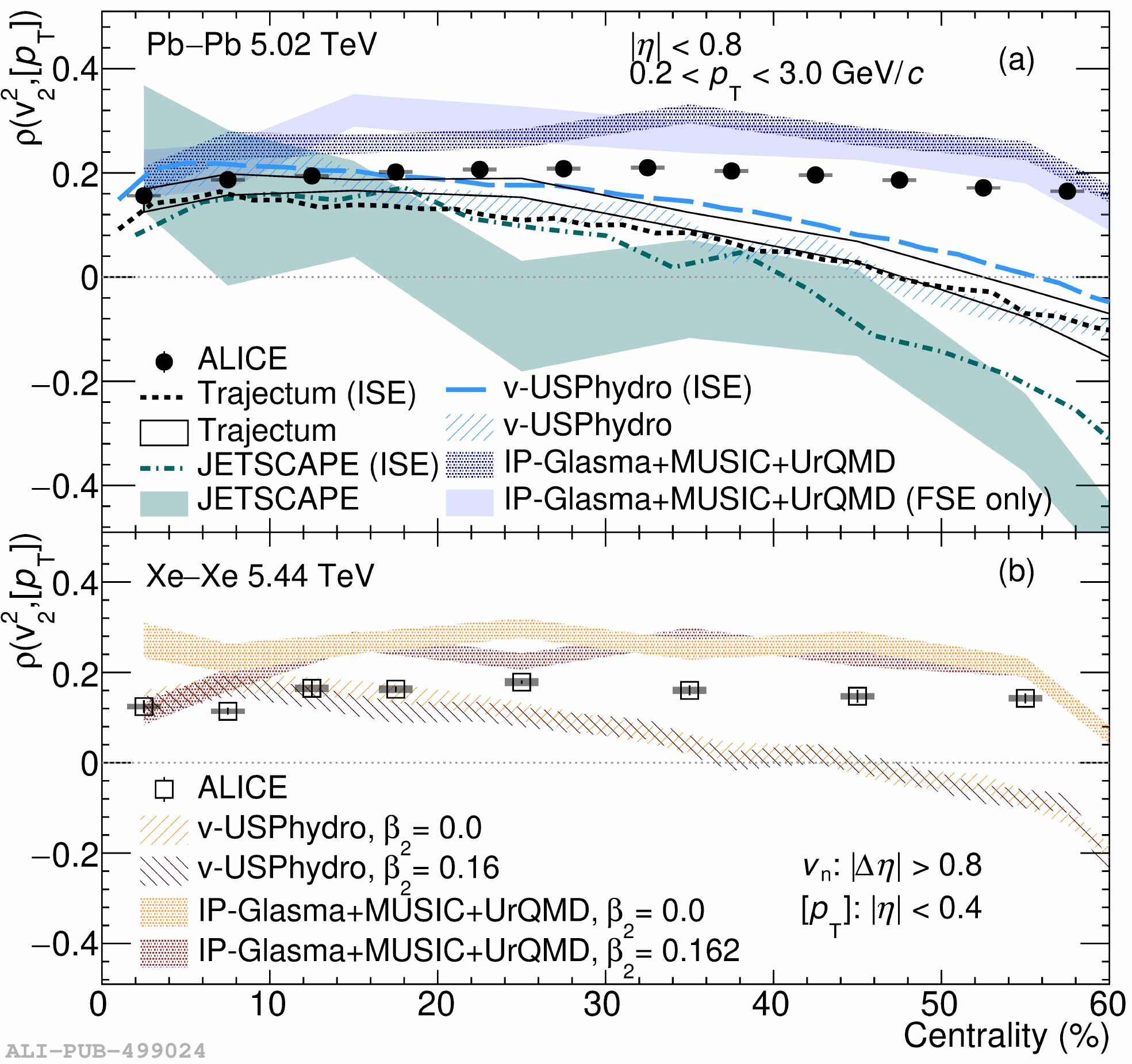}
 \end{center}
 \caption{ Centrality dependence of $\rho (v^2_n, [p_T]) $ in \PbPb  and \XeXe collisions at \fivenn  and \fivennFour, respectively, compared to different model calculations.}
 \label{fig:rhov2}
\end{figure}

Correlations of $v_2$ vs mean \pT, $\rho (v^2_n, [p_T]) $, as a function of centrality (Fig. \ref{fig:rhov2}) are mainly driven by the correlations of 
\noindent
the size and the shape of the system in the initial state,  and as such 
provide  a novel way to characterize the initial state and 
help pin down the uncertainty of the extracted properties of the quark--gluon plasma.
The sensitivity of the correlations of $v_2$ vs mean \pT vs centrality, $\rho (v^2_n, [p_T])$ to the initial conditions can be observed in Fig.~\ref{fig:rhov2} \cite{ALICE:2022xhd}. 
Data for \PbPb and \XeXe collisions are closer  to IP-Glasma initial condition as compared to Trento initial conditions. Therefore, including measurements of $\rho (v^2_n, [p_T]) $   vs 
centrality in Bayesian global fitting approaches could result in better constraint on the initial state in nuclear collisions.

\begin{figure}[H]
\begin{center}
 \includegraphics[width=0.95\linewidth]{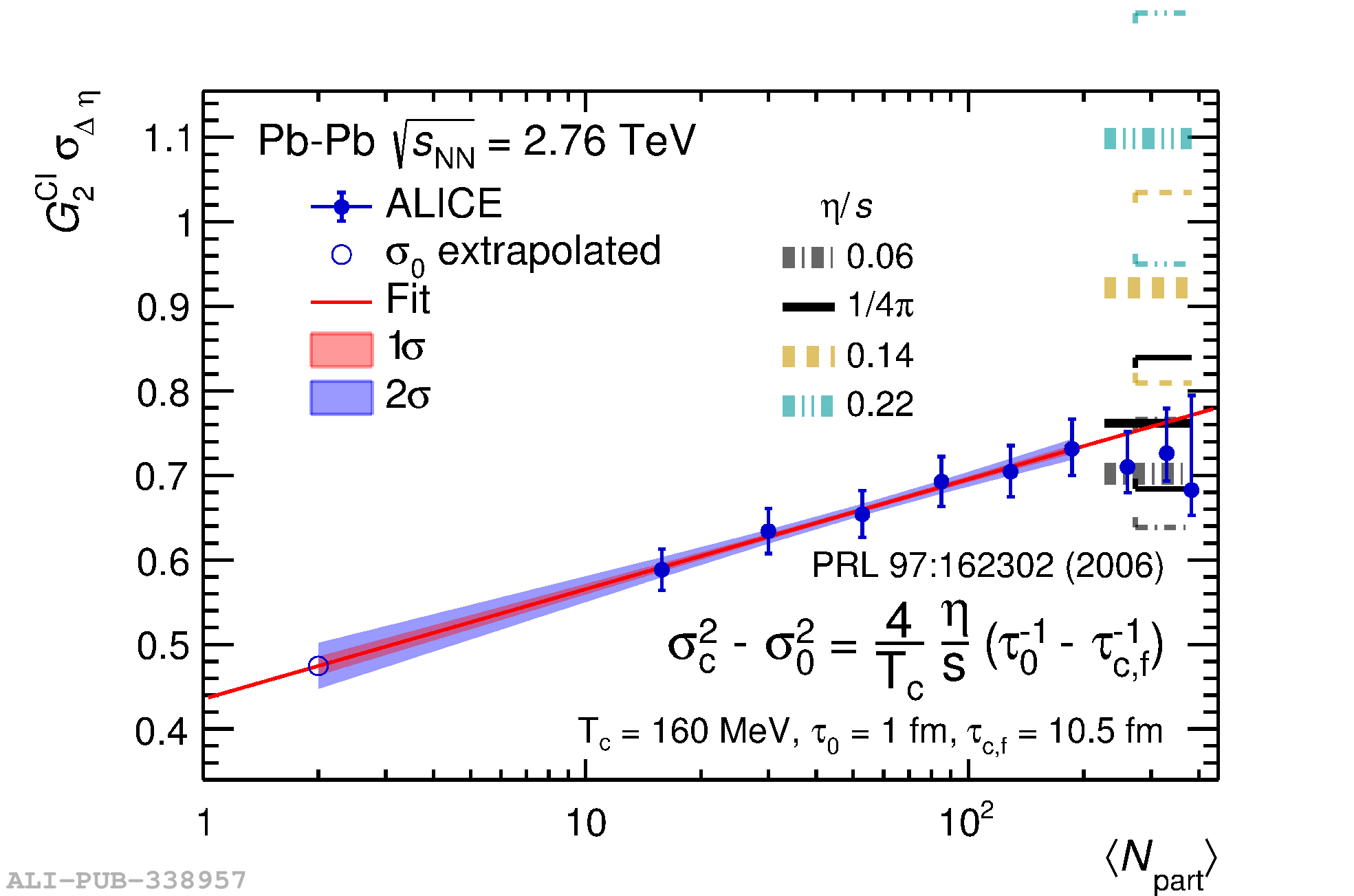}
\end{center}
 \caption{Width of $G_2$ in $\Delta \eta$ as a function of $\langle N_{\rm part} \rangle$ for \PbPb collisions at \twosevensixnn.}
 \label{fig:simaG2deltaeta}
\end{figure}

\begin{figure}[H]
\includegraphics[width=0.89\linewidth,trim={3cm 9cm 2.5cm 9cm }, clip]{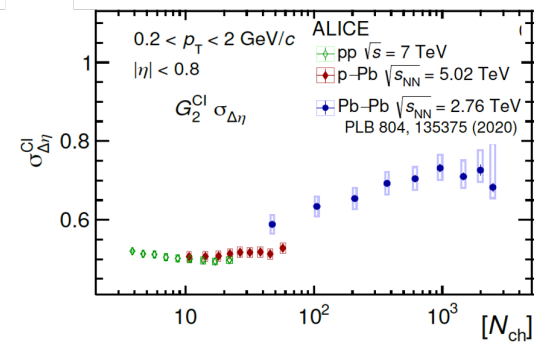}
  \caption{Width of $G_2$ in $\Delta \eta$ as a function of charge particle multiplicity for \pp, \pPb and \PbPb collisions.}
 \label{fig:simaG2deltaetaall}
\end{figure}

The two-particle transverse momentum correlator $G_2$ was proposed \cite{Gavin:2006xd,Sharma:2008qr} because of its sensitivity to the transport characteristics of the QGP.
The longitudinal width evolution with collision centrality carries information about  $\eta/s$ \cite{STAR:2011iio}, 
while it does not have any explicit dependence on the initial state fluctuations in the transverse plane of the system.  
The $G_2$ width evolution measured as a function of centrality in \PbPb collisions  at \twosevensixnn \cite{ALICE:2019smr} (Fig.~\ref{fig:simaG2deltaeta})  favours small values of $\eta/s$ 
at LHC energies. Furthermore, no evidence for shear viscous effects was found in \pp and \pPb collisions based on $G_2$ (Fig.~\ref{fig:simaG2deltaetaall}), or the system lifetime 
was too short for viscous forces to play a significant role \cite{ALICE:2022hor}.  
\noindent
ALICE data on the longitudinal width of $G_2$ were used  to compute values of \e/$s$ as a function of the charged particle multiplicity, obtaining $\eta/s$ values in the range 
from 0.04~$\pm$~0.02~(sys)  to 0.07~$\pm$~0.03(sys)  for LHC energies \cite{Gonzalez:2020bqm}.

\section{Electromagnetic radiation}
Photons and dileptons (lepton-antilepton pairs from internal conversion of virtual photons) are produced throughout all stages of the collisions. 
They are unique probes of the QGP because they leave the medium unaffected by the strong interaction. They provide information about the temperature and the radial expansion velocity of the QGP.

\begin{figure}[H]
\begin{center}
 \includegraphics[width=0.9\linewidth]{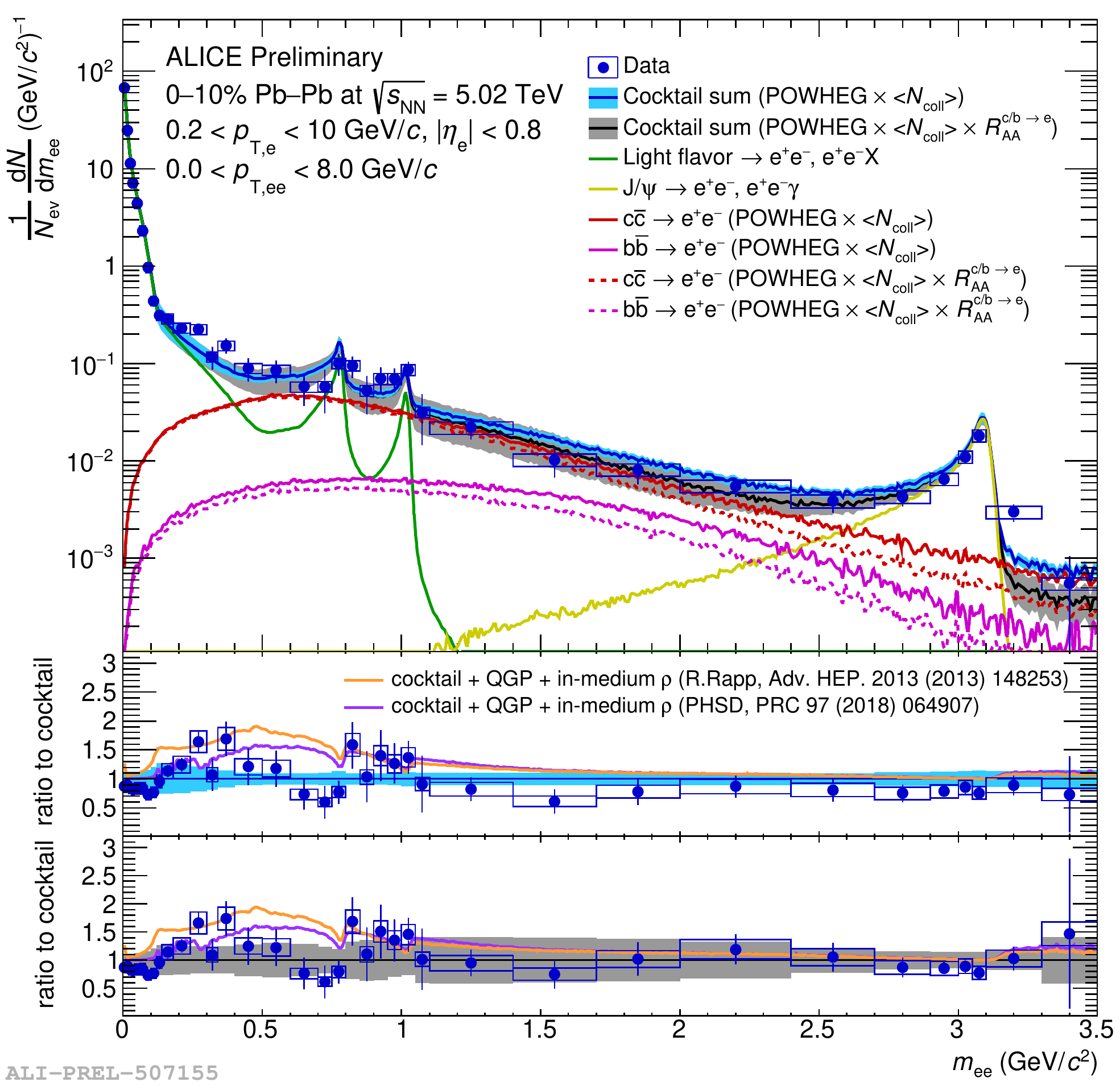}
\end{center}
\caption{Dielectron invariant mass  distribution for central 0-10\% \PbPb collisions at \fivenn compared  to the hadronic cocktail for two different versions of the correlated background from heavy flavour 
hadron decays. Middle and bottom plots show the respective ratio to the cocktails excluding the vacuum $\rho$. }
 \label{fig:dielectrons}
\end{figure}

Dileptons carry mass that can be used to distinguish between different sources of the radiation, and which allows for temperature measurements without a blue shift.
At low invariant mass (\mee $<$1.1 \GeVmass), the dielectron production is sensitive to the properties of short-lived vector mesons in the medium related to chiral symmetry 
restoration ($\rho$ -$a_1$, vector-axial vector meson mixing). 
The intermediate-mass region (1.2 \GeVmass  $<$ \mee $<$ 2.7 \GeVmass), is where the thermal black-body radiation dominates and the early temperature of the system can be extracted after subtraction 
of the very large background of correlated dielectron pairs from semi-leptonic charm and beauty hadron decays. 
Fig.~\ref{fig:dielectrons} shows the dielectron invariant mass  distribution for central 0-10\% \PbPb collisions at \fivenn compared  to the hadronic cocktail for two different versions of the correlated background from heavy flavour 
hadron decays. Middle and bottom plots show the respective ratio to the cocktails excluding the vacuum $\rho$. 
\noindent
Data show  an indication for an excess at lower mass compatible with the thermal radiation from the partonic and hadronic gas phase \cite{Rapp:2013nxa,Song:2018xca} up to \mee~$<$~0.5~\GeVmass while the model calculations overestimate 
the data for 0.5~$<$~\mee~$<$~0.7~\GeVmass. The intermediate mass region agrees better with the expectations of the models for the cocktail including HF suppression \cite{ALICE:2019nuy}.

\begin{figure}[H]
\begin{center}
  \includegraphics[width=0.9\linewidth]{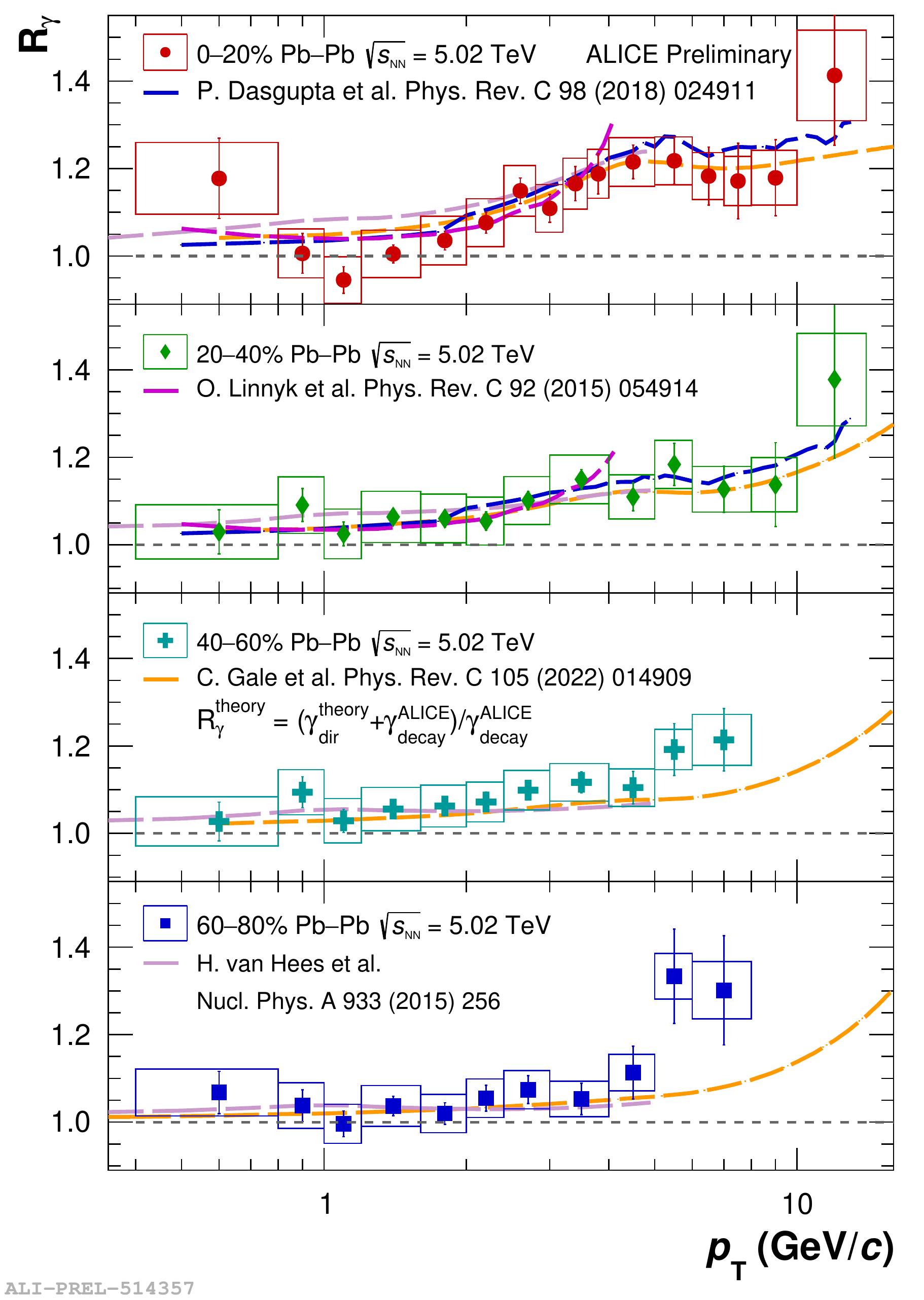}
\end{center}
 \caption{\Rg as a function of \pT from central (top) to peripheral (bottom) \PbPb collisions at \fivenn together with expectations from model calculations that include thermal photons, (and pre-equlibium photon)  and prompt photons.}
 \label{fig:Rgamma}
\end{figure}

Direct photons are photons not coming from hadronic decays. Experimentally, they can be measured using a  double ratio \Rg defined as 
\begin{equation}
R_\gamma = \frac{(N_{\gamma, inc}/N_{\pi^0})_{measured} }{(N_{\gamma,decay}/N_{\pi^0})_{sim}}
\end{equation}
in order to cancel correlated uncertainties. 
The decay photon spectra are estimated employing the decay cocktail framework \cite{Acharya:2018dqe}  using as input the measured \piz and \e meson spectra and \mT scaling for
non measured spectra.
The \Rg as a function of \pT is shown in Fig. \ref{fig:Rgamma} from central (top) to peripheral (bottom) \PbPb collisions at \fivenn. 
At low \pT where the thermal radiation dominates the measurement the value of \Rg 
is close to 1, which entails  a small  contribution from thermal and pre-equilibrium photons. For \pT~$>$~3~\GeVc, a value of  \Rg~$>$1 is measured which is attributed to prompt photons from hard scatterings.
At low \pT, model calculations with thermal and pre-equilibrium photons agree better with the data than if only prompt photons \cite{Gale:2021emg} are included.
Several model calculations that contain different assumptions (microscopic transport approach or relativistic hydrodynamic with different initial 
conditions, thermalization times with and without pre-equilibrium photons ) are able to describe the data. 
The current uncertainties do not allow for discrimination among the different theoretical model calculations \cite{Gale:2021emg,vanHees:2014ida,Paquet:2015lta,Chatterjee:2012dn,Chatterjee:2009rs,Linnyk:2015tha,He:2011zx
,Holt:2015cda,Heffernan:2014mla,Turbide:2003si}.  More precise results are expected with the full Run 2  and Run 3 data.

\section{Quarkonia}

\begin{figure}[H]
\begin{center}
 \includegraphics[width=0.87\linewidth,trim={1cm 6.6cm 1.cm 7cm}, clip]{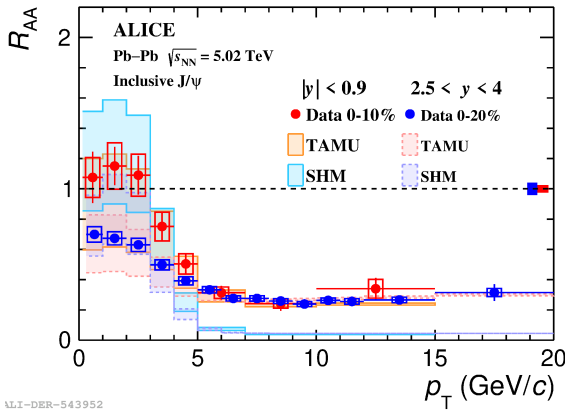}
\end{center}
 \caption{\RAA for $J/\psi$ as a function of \pT  for the central and forward rapidity regions in \PbPb collisions at \fivenn  compared to TH-TAMU and SHM model calculations.}
 \label{fig:Jpsi}
\end{figure}

Quarkonia are flavorless mesons  whose constituents are a heavy quark and its own antiquark ($c\bar{c}$ or $b\bar{b}$ for charmonium or bottonomium, respectively). Most of the states are stable with respect to strong 
decay into open charm or open bottom because of the mass threshold.
In the late 80s, it was realised that colour screening prevents $c\bar{c}$ binding in a deconfined medium \cite{Matsui:1986dk}. Consequently,  the observation of $J/\psi$ suppression 
would provide a signature of QGP formation. Furthermore,  the sequential suppression of the different quarkonia states would deliver information about the $T$ of the QGP
\cite{ALICE:2023gco,Andronic:2019wva,Du:2015wha}.  
Later, it was realised that another mechanism, the (re)generation, will compensate the suppression once the charm quark multiplicity would become  sizeable  \cite{Braun-Munzinger:2000csl,Thews:2000rj,Braun-Munzinger:2007edi}. 
The measured $J/\psi$ \RAA  as a function of \pT  for the central and forward rapidity regions in central \PbPb collisions at \fivenn  is displayed in Fig.~\ref{fig:Jpsi} compared to the TH-TAMU and SHM model calculations \cite{ALICE:2023gco}.
\RAA is higher at midrapidity than at forward rapidity for \pT $<$ 3 GeV/c in the most central collisions as can be explained by a large contribution from (re)generation to the $J/\psi$ yields.

\begin{figure}[H]
\begin{center}
 \includegraphics[width=0.85\linewidth]{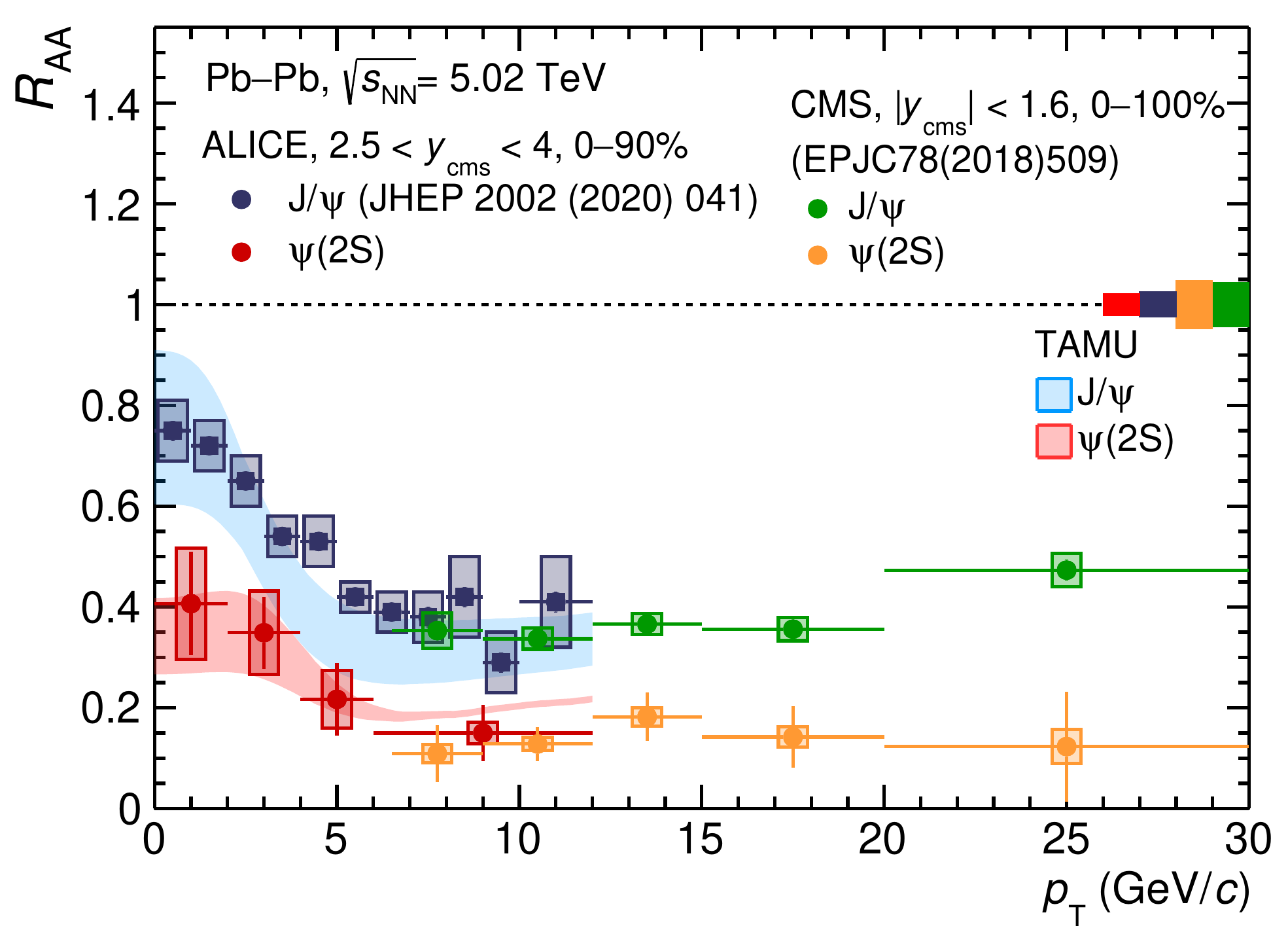}
\end{center}
 \caption{\RAA for $\Psi(2S)$  and  $J/\psi$ as a function of \pT  in \PbPb collisions at \fivenn compared  with theoretical model calculations and results from the CMS experiment.}
 \label{fig:Psi2s}
\end{figure}

The first accurate measurement of the inclusive $\Psi$(2S)  down to zero \pT  for 2.4$<y<$ 4 has been achieved by ALICE  \cite{ ALICE:2022jeh}. 
Fig.~\ref{fig:Psi2s}  displays  the \pT-dependent $\Psi$(2S) \RAA compared to the $J/\psi$ \RAA.
The $\Psi(2S)$ has ten times lower binding energy and shows two times larger suppression at low \pT as compared to the $J/\psi$. 
The strong increase of  \RAA towards low \pT  for both quarkonia states is an indication of (re)generation.
The ALICE results are in good agreement with  CMS results and with theoretical model calculations.

\section{Open heavy flavour}

 Heavy quarks are interesting probes to characterize the QGP, because due to their large mass the production even at low \pT is driven by hard scatterings in the early stages of the collision 
and additional thermal production is negligible. 
\begin{figure}[H]
\begin{center}
 \includegraphics[width=0.9\linewidth,trim={0 0 10cm 0 }, clip]{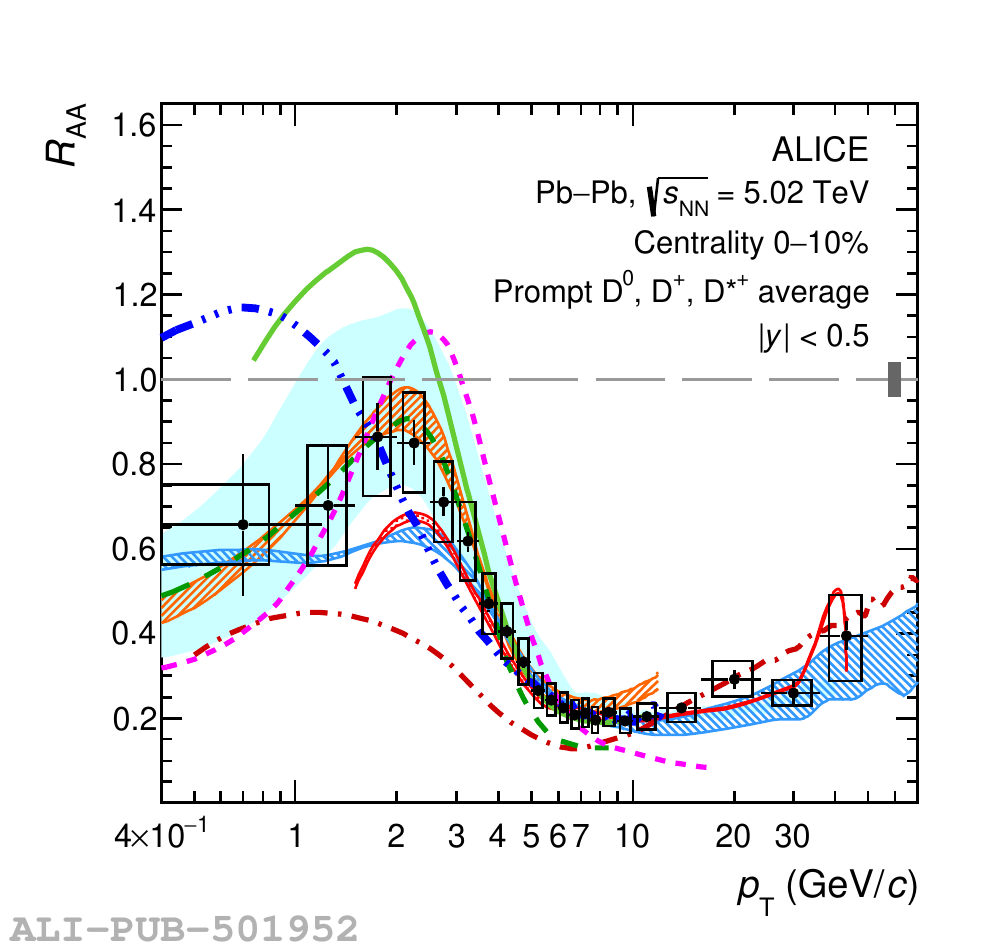}
\end{center}
 \caption{Average \RAA of prompt D$^0$, D$^+$, and D$^{\star+}$ mesons  in the 0--10\% centrality class in \PbPb collisions at \fivenn compared with pre\-dic\-tions 
 of mo\-dels im\-ple\-men\-ting charm-quark transport in a hy\-dro\-dynami\-cally expanding medium.}
 \label{fig:D0}
\end{figure}
\noindent
Thus, heavy quarks experience the full space-time evolution of the hot and dense QCD medium. At low \pT, heavy-quarks undergo Brownian motion in the medium, 
carrying, hence, information of the equilibration process. High \pT heavy quarks interact with the QGP and lose energy  
via  inelastic (radiative) processes. Open heavy flavor can be measured in ALICE through various channels. Some hadronic decay channels can be fully reconstructed to zero \pT.
\begin{figure}[H]
\begin{center}
 \includegraphics[width=0.9\linewidth,trim={10cm 0 0 0 }, clip]{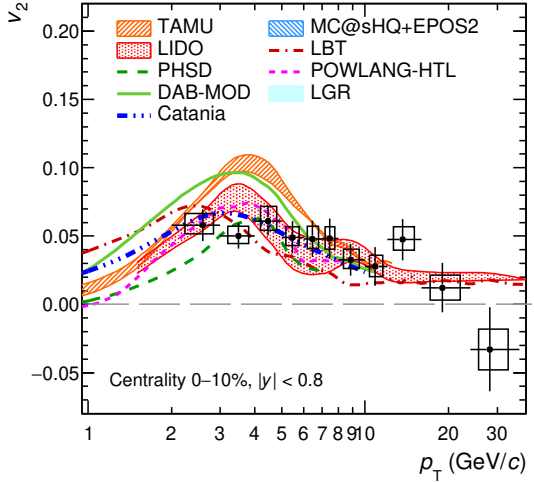}
\end{center}
 \caption{Average elliptic flow $v_2$   of prompt D$^0$, D$^+$, and D$^{\star+}$ mesons  in the 0--10\% centrality class in \PbPb collisions at \fivenn compared with predictions 
 of models implementing the charm-quark transport in a hydrodynamically expanding medium.}
 \label{fig:D0v2}
\end{figure}
\begin{figure}[H]
\begin{center}
 \includegraphics[width=0.9\linewidth]{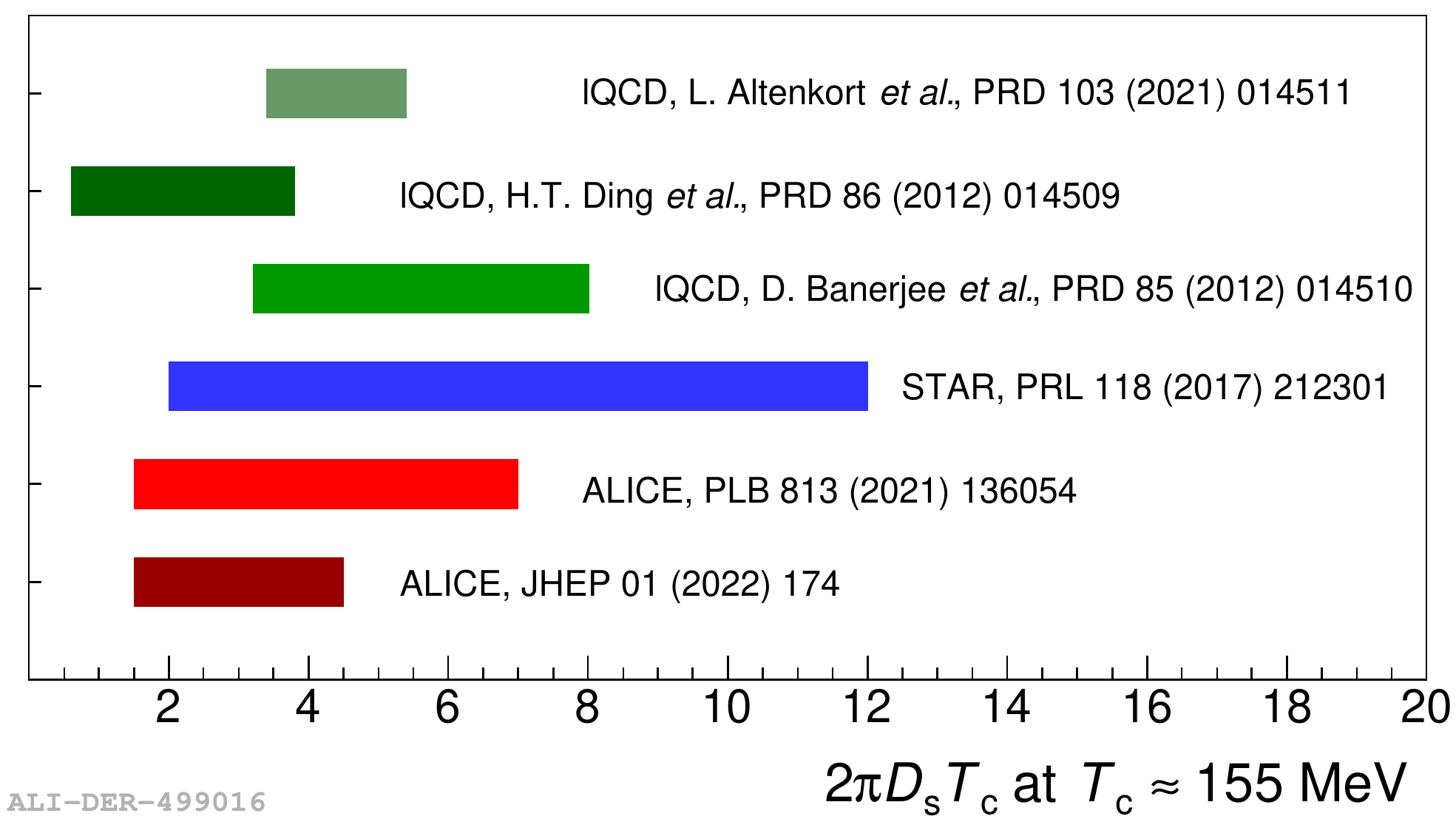}
\end{center}
 \caption{Compilation of values for the  spatial diffusion coefficient $D_s$ with their respective  uncertainties as obtained by measurements and model calculations.}
  \label{fig:D0ranges}
\end{figure}

The  average \RAA  of prompt D$^0$, D$^+$, and D$^{\star+}$ mesons  for \PbPb collisions in the 0--10\% centrality class  is shown in Fig.~\ref{fig:D0} \cite{ALICE:2021rxa}.
 Compared to the \RAA of charged pions it is larger for \pT $<$ 8 \GeVc. All models shown in  Fig.~\ref{fig:D0}  agree with the data, with some tension at low \pT.  
 The comparison of D-meson \RAA and $v_2$ (Fig.~\ref{fig:D0v2} ) to models implementing charm-quark transport in a hydrodynamically expanding medium gives information of the 
 interaction of heavy quarks with the medium, constraining the  spatial diffusion coefficient. 
 The resulting spatial diffusion coefficient $D_{\rm s}$  extracted from the ALICE measurement (1.5$< 2\pi D_{\rm s} T_{\rm c} <$4.5) together with a compilation of the values as obtained from measurements and 
 model calculations is shown in Fig.~\ref{fig:D0ranges}. This value of $D_{\rm s}$  translates into a relaxation time of the charm quark $\tau_{\rm charm}\sim$~3-8~fm/$c$.

\section{Jets}
The production of jets, i.e. collimated sprays of particles arising from fragmentation of partons produced in high $Q^2$ interactions, in \pp collisions can be well described
in pQCD calculations. Thus, by measuring jet suppression in \PbPb  collisions over a wide set of parameters, properties of the QGP can be extracted. 
\begin{figure}[H]
\begin{center}
 \includegraphics[width=0.9\linewidth]{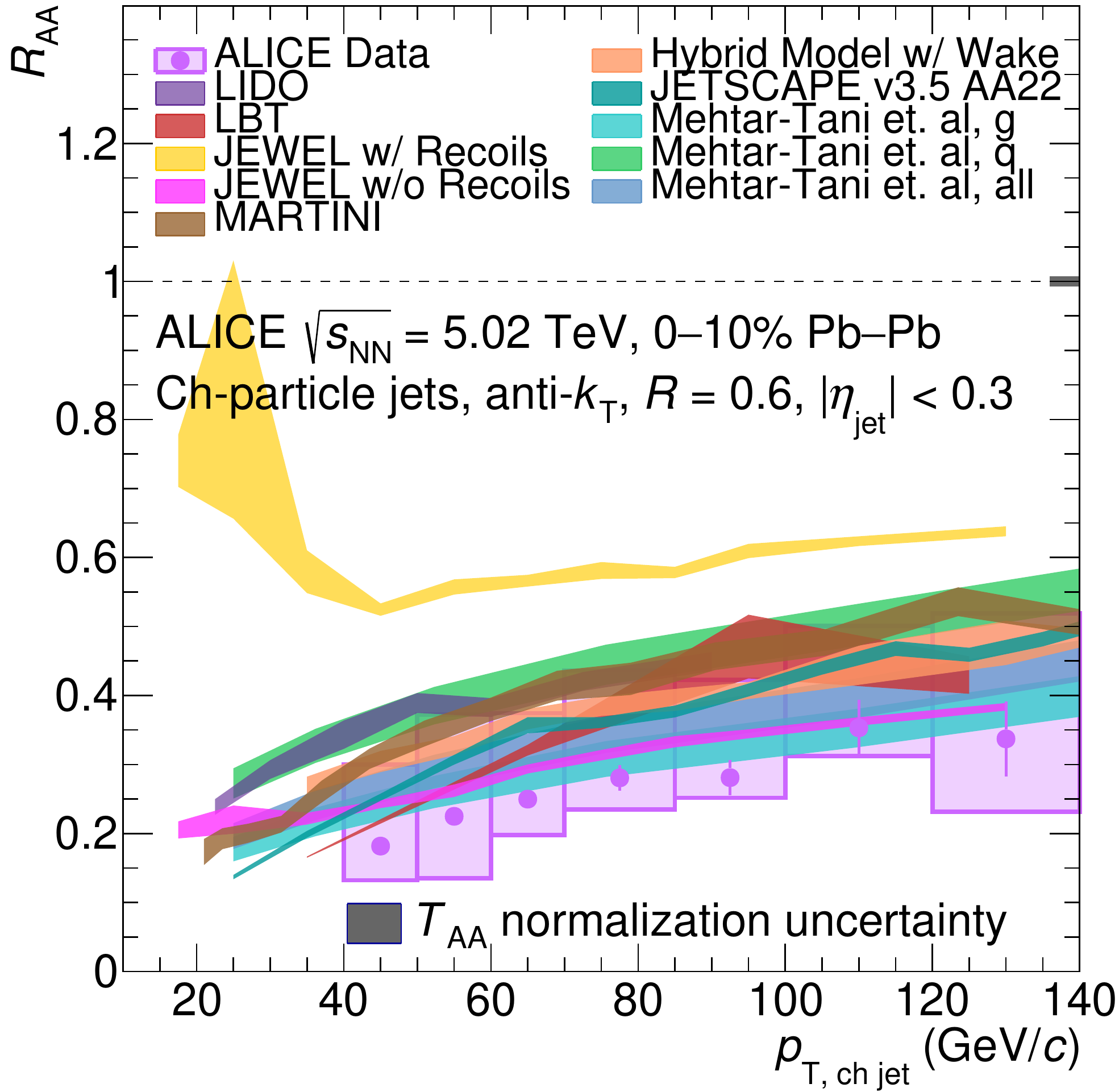}
 \end{center}
 \caption{Nuclear modification factor (\RAA) for charged particle jets  with R= 0.6  obtained using the ML-based method compared to  different model predictions \cite{ALICE:2023waz}.}
  \label{fig:Jets}
\end{figure}
Jet suppression is quantified using the \RAA with a novel  machine learning  (ML) based approach to subtract underlying Pb-Pb event fluctuations from jet energy (see Fig.\ref{fig:Jets}) that allows
measurements at transverse momenta as low as  40~\GeVc for a large resolution parameter of $R=$~0.6 in most central \PbPb collisions (or \pT~$>$~20 \GeVc for $R=$~0.2) \cite{ALICE:2023waz}.
A large suppression is still present  for $R=$~0.6  implying that the lost energy is still not recovered within the jet "cone". A comparison of \RAA for $R=$~0.6 and $R=$~0.2 
indicates that the suppression may be even larger for large-cone ($R=$~0.6) low-\pT jets.
The larger acceptance achieved both in \pT and resolution parameter $R$, thanks to the ML-based approach, is crucial for discriminating among different models.
A comparison of model calculations to the jet \RAA is shown in Fig.~\ref{fig:Jets}).
The calculations generally describe the data in central collisions for the smaller resolution parameters ($R=$~0.2 and 0.4). JEWEL with recoils predicts significantly higher values for \RAA at $R =$~0.6 than measured.

\section{ALICE in Run 3  and Run 4 at the LHC}

ALICE went through a major upgrade during the LS2 which enables data taking in a  continuos readout mode to profit from the increased LHC luminosity and  improves the tracking resolution \cite{Citron:2018lsq}.   
The upgrades comprise a new inner tracking system (ITS2), new GEM-based readout chambers of the time projection chamber, new muon
forward tracker (MFT), new trigger and readout and a new online/offline (O2) software \cite{ALICE:2023udb}.

\begin{figure}[H]
\begin{center}
 \includegraphics[width=0.8\linewidth]{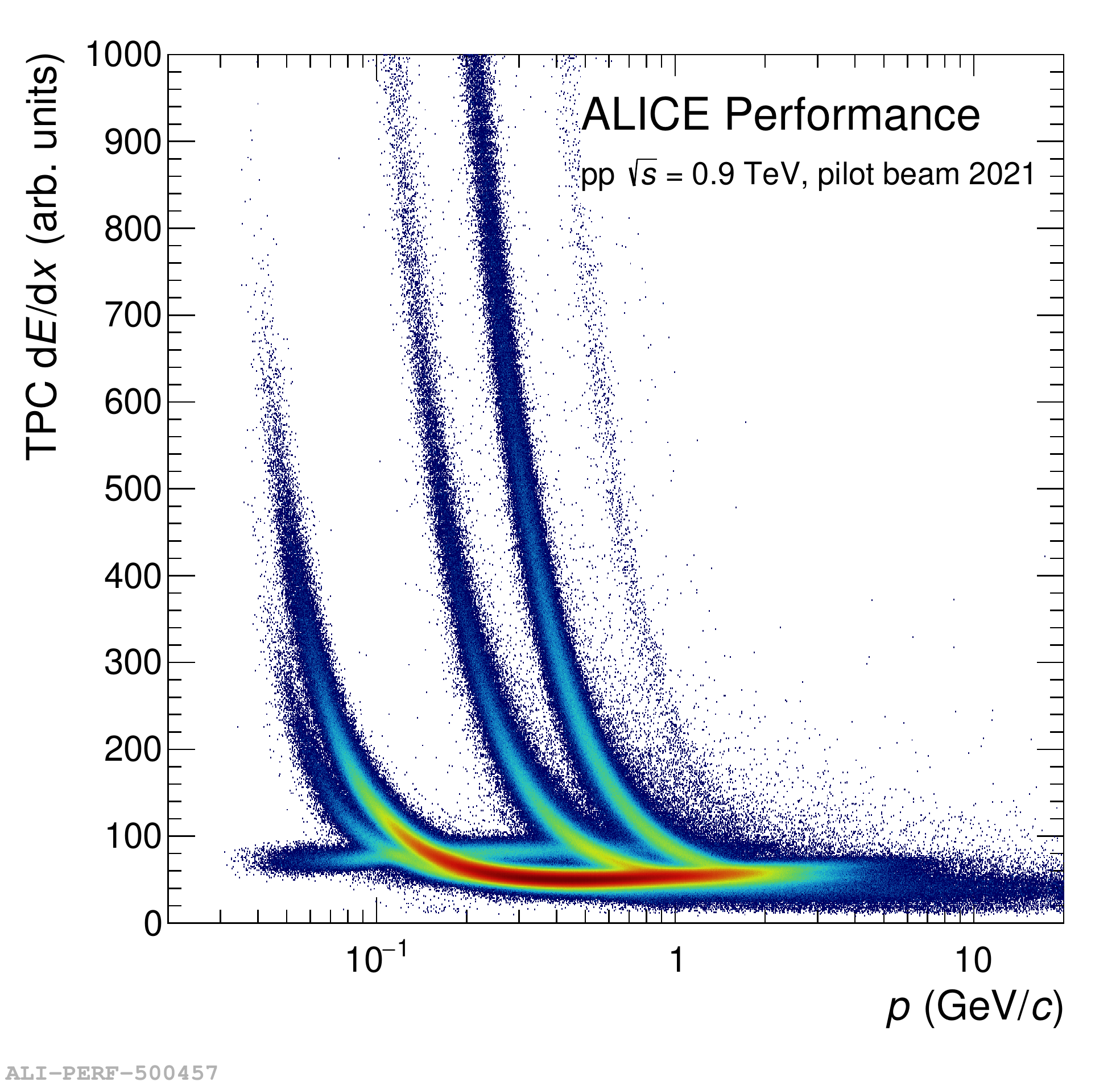}
\end{center}
 \caption{Specific energy loss (d$E$/d$x$) in the TPC versus momentum for \pp collisions at \nineH during the 2021 pilot beam.}
 \label{fig:PIDRun3}
\end{figure}

On July 5, 2022,  LHC Run 3 started officially: proton beams collide now at a center of mass energy of 13.6 TeV, the highest energy ever reached. 
Commissioning with pilot beams during 2021 showed the expected performance of the upgraded experiment in terms of tracking and particle identification (see Fig. \ref{fig:PIDRun3}).  
A total integrated luminosity of 15 $pb^{-1}$ was recorded during 2022 at different interaction rates ranging from 3 kHz up to 1 MHz. Additionally, a test run with Pb beams at \fivennThree
 took place prior to the winter break that is crucial to prepare and optimize the data processing and compression. 
During the upcoming LS3 (see Fig.~\ref{fig:LHCTimeLine}),  the FoCal detector \cite{CERN-LHCC-2020-009} will be installed and the current ITS2 will be replaced  by the ITS3 \cite{Musa:2703140}.

\section{ALICE 3 for Run 5 and Run 6}

A huge progress on the characterization of the QGP was already achieved thanks to ALICE results in Run~1 and Run~2,  and will continue to deepen  with the Run~3 and Run~4 scientific program.
However, several essential  questions like the fundamental properties of the quark-gluon plasma driving its constituents to equilibration, 
a comprehensive study of the QGP hadronization mechanisms, the partonic equation of state and its temperature dependence and the
underlying dynamics of chiral symmetry restoration will still remain unanswered after Run~3 and Run~4.
Thus, the ALICE Collaboration  proposed to install  a next generation multipurpose detector, ALICE 3 \cite{Adamova:2019vkf,ALICE:2022wwr}, for Run 5 and Run 6 (see Fig.~\ref{fig:LHCTimeLine}). 
\begin{figure}[H]
\begin{center}
\includegraphics[width=0.67\linewidth,angle=-90,trim={1.5cm 2cm 1.5cm 2cm }, clip]{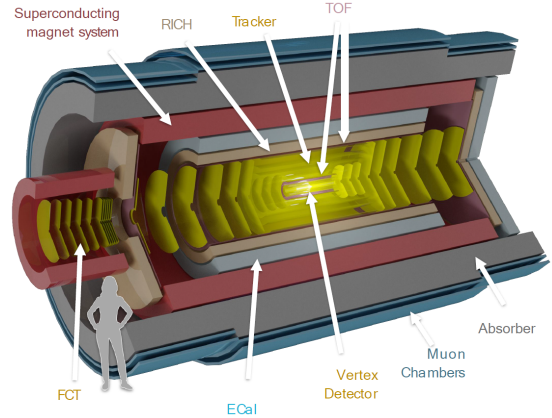}
\end{center}
 \caption{Schematic view of the ALICE 3 experiment with the inner and outer trackers, RICH, TOF detectors, ECAL and muon detectors in the central and foward/backward barrels, and the forward conversion tracker (FCT) specialized in low momentum photons.}
 \label{fig:ALICE3}
\end{figure}
\noindent
ALICE 3 (see Fig.~\ref{fig:ALICE3}) is based on the use of monolithic active pixel sensors (MAPS) in combination with deep submicron commercial CMOS technologies.
The rate capabilities would be a factor of about 50 higher with respect to ALICE in Run 4, being able to
exploit the whole delivered p-A and A--A luminosities. ALICE 3 is a large acceptance detector, covering a pseudorapidity range of $|\eta|<$~4   and  \pT~$>$~0.02 \GeVc. 
The expected track point resolution  is $\sigma_{DCA} \sim $ 10 $\mu$m at a \pT~$=$~0.2 ~\GeVc. The detector will deliver particle identification for $\gamma$, $e^\pm$, $\mu^\pm$, $K^\pm$ and  $\pi^\pm$.
Such an experiment will be able to deliver  systematic measurements of 
(multi)heavy-flavoured~hadrons down to low \pT, precision differential measurements of dileptons complemented with direct photon measurements, 
as well as correlation and fluctuation measurements over a large rapidity range among others. 

\begin{figure}[H]
\begin{center}
 \includegraphics[width=0.85\linewidth]{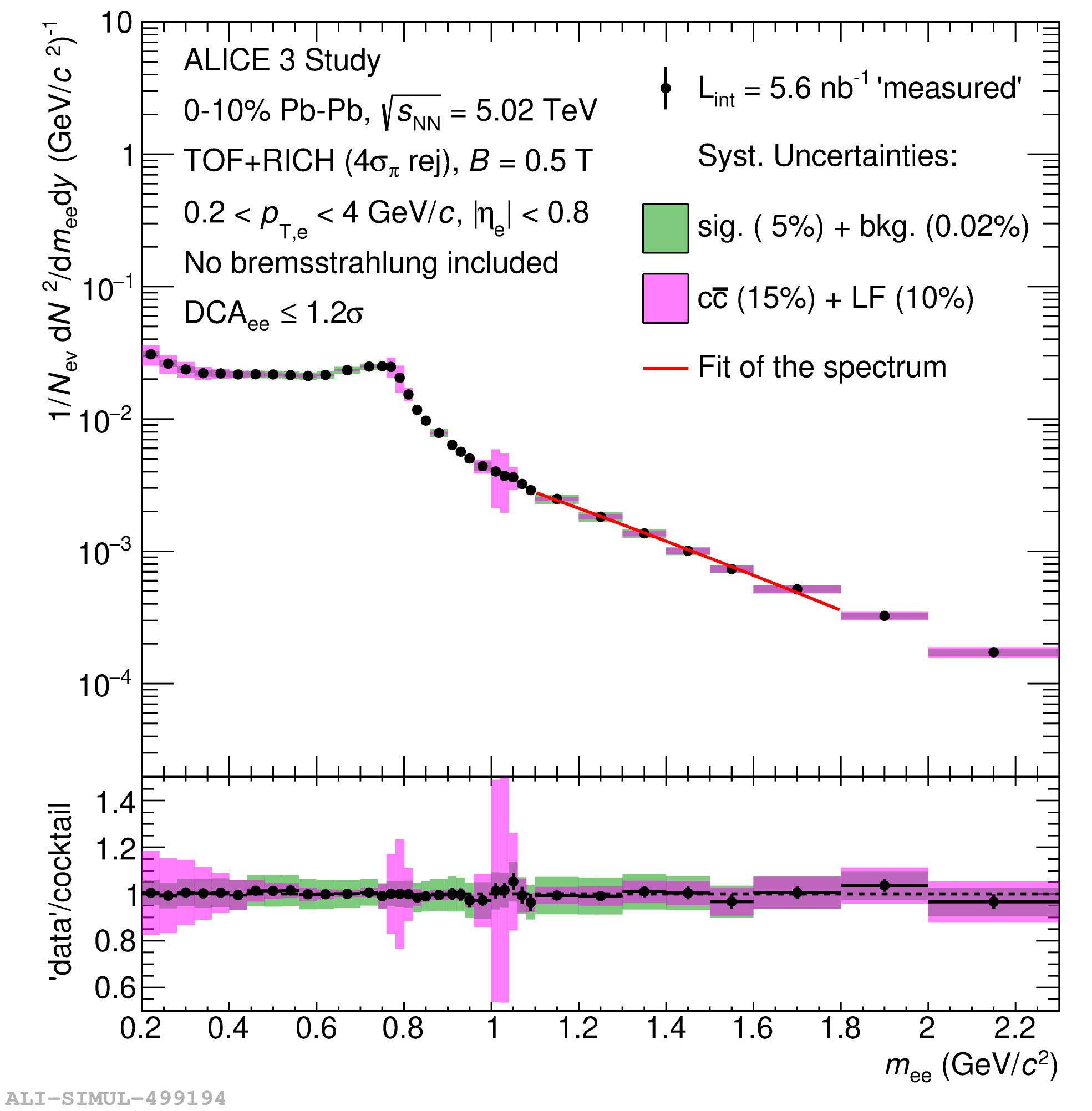}
\end{center}
\caption{Dielectron invariant mass spectrum ($m_{ee}$) fitted with Eq. \ref{eq:Tmee} for $m_{ee}$ $>$ 1.1 \GeVmass. }
\label{fig:highmass}
\end{figure}

 The double-differential analysis of dilepton production in transverse momentum and mass provides access to the time evolution of the temperature.
The excellent capabilities  of ALICE 3 in terms of good $e$-PID down to low \pT, small detector material budget ($\gamma$ background)
and an excellent pointing resolution (heavy-flavour decay electrons), translate into the expected invariant mass spectrum of thermal $e^+e^-$ pairs produced in central \PbPb collisions at 
\sNN~=~5.02~TeV as depicted in Fig.~\ref{fig:highmass}.
It would be possible to probe the time dependence  of the QGP  temperature by  fitting the thermal dielectron \mee spectrum for $m_{ee}$ $>$1.1 \GeVmass (QGP radiation dominated) in 
bins on pair \pT,ee with 
\begin{equation}
\frac{dN_{ee}}{dm_{ee}} \propto  (m_{ee}T)^{3/2}e^{-m_{ee}T}.
\label{eq:Tmee}
\end{equation}
The thermal dielectron elliptic flow as a function of \mee and \pT,ee would be within reach with ALICE 3.  
Furthermore, the high precision thermal dielectron spectrum at $m_{ee} < $1.2 \GeVmass  allows to access the chiral symmetry restoration effects like the $\rho$-$a_1$ mixing \cite{Rapp:2013nxa,Hohler:2013eba,ALICE:2022wwr}.
\noindent
 Real photons will be measured  in ALICE 3 with the photon conversion method  (PCM) and with the electromagnetic calorimeter (ECAL). 
 The excellent tracking of photon conversion products together with the large acceptance will allow high precision measurements of \piz and \e down to zero \pT  reducing the systematic uncertainties 
 of the dominant photon background sources and, therefore, of the direct photon measurements.
 ALICE 3 will then improve on the measurements of the temperature and elliptic flow of direct photons.
By combining the information of  real and virtual photons the temperature and the radial expansion  velocity of the QGP can be determined.

 \section{Conclusions}
Much progress on the characterization of the QGP was achieved thanks to  ALICE  results in LHC Run~1 and Run~2. Some of the questions
that remained open mainly due to statistics will be addressed during the already ongoing  Run~3 and Run~4. 
Others will have to be postponed to Run~5 and Run~6 with the completely new ALICE 3 experiment.

\noindent

\vspace{1cm}

\noindent
{\large \bf Acknowledgements}\\

\noindent
The author would like to thank the organizers for the invitation to present an overview of the ALICE results. 
I am very grateful for the fruitful discussions and  warm atmosphere during the workshop.

\end{multicols}

\medline
\begin{multicols}{2}
%
\nocite{*}
\bibliographystyle{rmf-style}
\bibliography{ref}
\end{multicols}
\end{document}